\documentclass[preprint,authoryear]{elsarticle}
\journal{Journal of Informetrics}

\usepackage{graphicx}
\usepackage[colorlinks=true]{hyperref}

\begin{document}
\title{Bibliometric author evaluation through linear regression on the coauthor
network}
\author{Rasmus A. X. Persson}
\ead{rasmus.a.persson@gmail.com}
\address{Department of Chemistry \& Molecular Biology, University of
Gothenburg, SE-412 96 Gothenburg, Sweden}

\begin{abstract}
The rising trend of coauthored academic works obscures the credit assignment
that is the basis for decisions of funding and career advancements. In this
paper, a simple model based on the assumption of an unvarying ``author
ability'' is introduced. With this assumption, the weight of author
contributions to a body of coauthored work can be statistically estimated. The
method is tested on a set of some more than five-hundred authors in a coauthor
network from the CiteSeerX database. The ranking obtained agrees fairly well
with that given by total fractional citation counts for an author, but
noticeable differences exist.
\end{abstract}


\maketitle

\section{Introduction}
Typical quantitative indicators of scientific productivity and quality that
have been proposed---be it on the level of individuals, institutions or even
whole geographic regions---are, in some form or another, ultimately based on
the citation distribution to previous (and available) scientific works (in this
paper referred to as ``papers'' for short for all types [books, regular
articles, rapid communications, commentaries, proceedings, \textit{etc.}]). A
fairly extensive scientific literature exists on the subject of discriminating
between individuals or scientific institutions, motivated to a large extent by
the perceived need of the merit-based distribution of funding which is scarce
in relation to the number of active scientists. Such indicators range from the
simple (counting the number of papers and/or citations) to the more elaborate,
such as the $h$-index
\citep{hirsch05,jin06,hirsch07,bornmann05,bornmann07,bornmann08} and its many
variants
\citep{egghe06,kosmulski06,jin07a,jin07b,egghe08,bras-amoros11,ausloos15}. For
a recent and in-depth review of the fundamentals this topic (citation
counting), see the paper by \cite{waltman16}. This comparison is
in some schools of bibliometrics developed further in that the incoming
citations to a paper are weighted by the importance of the citing source. This
importance can be defined, for instance, from the number of citations the citing
paper has itself received, or the number of citations of the citing author. For
a review of this topic and an empirical investigation of its robustness, see
the paper by \cite{wang16}.

In this paper, we are motivated by the confounding factor that coauthorship
poses to any such analysis. Different options for dealing with this problem
have been proposed. The simplest is to divide the credit equally among all
contributing authors \citep{batista06,schreiber08} (known both as ``fractional
counting'' or ``normalized counting''); after that comes weighting author
credit by a simple function of the author's position in the author list
\citep{hagen09,sekercioglu08,zhang09}, or even more intricate schemes based on
this notion \citep{aziz13}. However, these alternatives cannot be motivated by
more than ``hunches'' about how a particular ``authorship culture'' assigns
credit. Clearly, a quantitative approach is more scientific than a qualitative,
or worse, arbitrary one.  Special mention is here given to the papers by
\cite{tol11}  and  by \cite{shen14}, in which
intuitive statistical models are used to disentangle the coauthorship
contributions.

Tol's \citeyearpar{tol11} idea may be summarized as follows. Whenever
two authors write a joint paper and it is highly cited, the senior
author of the pair\footnote{Defined in terms of ``Pareto weights'' which are
directly related to the average citations per article of an
author.} should receive a disproportionally large share of the citation credit.
The rationale for this is that it is more typical of the senior author, judging
from past experience, to write highly cited papers, and it is therefore
reasonable to assume that her contribution is more responsible for the ultimate
quality.  With his method and a limited sample set comprising some fifty
authors, \cite{tol11} finds small deviations of up to 25\% between
his ``Pareto weights'' and what he terms ``egalitarian weights'' in which
coauthorship credit is equally distributed.

\cite{shen14} agree with \cite{tol11} on
the principle of assigning more credit to the ``senior author'', but the
algorithm to determine the actual credit assignment is different. To determine
the ``relative seniority'' of each coauthor, their algorithm weighs both the
number of papers by the author \emph{and} the degree to which these papers
share citations from papers citing the one under consideration. In this way,
papers that are more ``similar'' to the one under consideration contribute more
to the ``seniority'' of that coauthor when assigning the authorship credit.

The idea behind the present paper is basically the same, but the execution is
different. Rather than assume a fixed form of a distribution like \cite{tol11}, we assume a fixed form for the underlying ``ability'' to
produce said distribution in the first place. We then solve for this ``author
ability'' statistically to find those authors who consistently manage to
contribute to ``high-quality'' papers. Another difference, which also
distinguishes the method from that by \cite{shen14},
is that a junior author is not necessarily ``punished'' for publishing with a
senior coauthor. If a paper is very successful compared to previous papers on
the topic, it is not altogether unreasonable to assume that this atypical
performance should be disproportionately credited to any authors not
participating in the earlier work. However, in both \cite{shen14} and in
\cite{tol11}, credit is instead disproportionately allocated to the senior
author. Much like \cite{tol11}, the rigorous application of our method requires
knowledge of complete coauthor networks, and can only be approximately applied
otherwise.  This is, however, more of a formal problem than a practical one.

\section{Regression model for coauthorship contribution}
We assume that the arbitrary author $i$ has an unchanging ability $a_i$ for
contributing to scientific papers.\footnote{This assumption does not contradict
the statement in the Introduction that ``a senior author, judging from past
experience,'' is more typically able to write highly cited papers. The senior
author may always have been good at producing highly cited scientific output,
but contrary to the case of the junior author, she has the credentials to back
it up.} A paper $\alpha$, once produced, possesses a
``scientific quality'' that we non-committally denote by $q_\alpha$ for now.
This variable could be, for instance, the total number of citations or the
rate of citation accumulation, to name a few. For notational simplicity, we
define the elements, $f_{\alpha i}$, of a dimensionless ``authorship tensor''
$\mathbf{F}$, to be unity if author $i$ contributes to paper $\alpha$, and zero
otherwise:
\begin{equation}
f_{\alpha i} = \left \{ \begin{array}{lr} 1, & \mathrm{if\ } i \mathrm{\ is\
author\ of\ } \alpha \\
0, & \mathrm{otherwise} \end{array} \right .
\end{equation}
With these definitions, we now define $a_i$ through,
\begin{equation}
\label{eq:quality}
\ln q_\alpha = \sum_{i=1}^{M_\mathrm a} f_{\alpha i} \ln a_i
\end{equation}
where $M_\mathrm a$ is the total number of authors in the statistical sample,
formally the number of individuals who have ever produced a work of science.
In practical calculations, we limit ourselves to much smaller subsets of
authors in a citation database. With modern computers, solving the complete
system of equations is possible if one has access to the entire database.
Typically, for individuals, the database is only partially accessible through
search keywords of an online interface and the database in its entirety is not
allowed (because of commercial contracts between the library and the database
provider, for instance) to be downloaded and mined for its data. Such a
limitation does not pose a greater problem than the reduction of the underlying
statistical data.

Before we continue, we note that the choice of the logarithm function in
Eq.~(\ref{eq:quality}) is judicious.  First, it implies that ``the whole is not equal to the sum of its parts'' and is meant to capture at least some of the
synergistic effects of a collaboration (as suggested, for instance, by
\cite{figg06}): in other words, the
relation between the number of authors and the resulting quality of the paper
is taken to be non-linear rather than linear. Here, we follow \cite{ke13}
closely, but replace his ``paper fitness'' by our ``author
ability''. Ke's model is more general, but we do not want to proliferate the
number of fitting parameters needlessly. Second, since the value of $q$ may
vary over several orders of magnitude in typical cases (\textit{vide infra}),
the logarithm ensures a more modest range for the regression. This said,
Eq.~(\ref{eq:quality}) is obviously an \textit{Ansatz} chosen merely for its
simple mathematical form rather than being based on some underlying physical
understanding of research production within collaborations.

If among themselves, $M_\mathrm a$ authors have published exactly
$M_\mathrm a$ papers, Eq.~(\ref{eq:quality}) forms a system of $M_\mathrm a$
linear equations that can be solved, in principle, for the unique set
$\{a_i\}_{i=1}^{M_\mathrm a}$ of author abilities if the determinant of the
square matrix
\begin{equation}
\mathbf{F} = \left [ \begin{array}{c c c} f_{11} & \cdots & f_{1 M_\mathrm a} \\
										\vdots & \ddots & \vdots \\
										f_{M_\mathrm a 1} & \cdots & 
										f_{M_\mathrm a M_\mathrm a}
										\end{array} \right ]
\end{equation} 
is non-zero. Such a situation is \textit{a priori} atypical, and the more
common case is where the number of papers, $M_\mathrm p$, does not equal
$M_\mathrm a$. However, the methods of statistical fitting (\textit{e.~g.},
least-squares) can still produce a set $\{a_i\}_{i=1}^{M_\mathrm a}$, which may
be unique or not depending on the circumstances. Hence, the proposed method may
be seen as the regression analysis for the unknown ``author ability''
underlying quality scientific paper production. The method of least squares is
the one which we will employ in this work. It has two desirable properties:
first, it is sensitive to outliers, and thus to very productive or skilled
researchers---a concern raised principally by Egghe in his $g$-index
\citep{egghe06}; second, it is numerically easier to handle than, say, the
least-absolute error. For clarity, we note that the error function which we
seek to minimize is the sum of the squared residuals:
\begin{equation}
R(\{a_i\}) \equiv \sum_\alpha \left (\ln q_\alpha - \sum_i f_{\alpha i}\ln a_i
\right )^2
\end{equation}

In a set of scientific papers, the quality---however defined---will exhibit a
distribution over the papers. The least-squares fitting of the set $\{\ln
a_i\}$ to the set $\{\ln q_\alpha\}$ may, if no further constraints are
present, lead to negative values in the former set. While this is reasonable
from a statistical point of view, it seems self-contradictory from a physical
point of view that the addition of an extra author to a paper may lead to a
decline in the quality of the resulting product. Therefore, in this
paper we always impose the extra condition $\ln a_i \geq 0$ for all $i$ in the
author set. The least-square solution of Eq.~(\ref{eq:quality}) may then be
found by, for instance, iterative gradient minimization techniques.

\subsection{On the interpretation of the meaning behind the author ability
variable}
From the purely mathematical perspective of author ranking, the condition that
$\ln a_i \geq 0$ is not strictly necessary and there would be some numerical
benefits for the solution of Eq.~(\ref{eq:quality}), were it
to be relaxed. For one thing, the residuals in the regression would be
decreased. However, we stick to this condition in this paper because we
want to maintain at least some ``physical'' connotation for the $a$-values. If
we allow negative values for $\ln a$ in the fitting, we basically say that
adding an author to a collaborative work may lead to a decrease in the
resulting quality. However---\emph{assuming the scientific field in which the
paper is produced is sufficiently rigorous to permit a general consensus of the
importance of results}---it should be clear that such a situation is only
possible if the coauthors allow the quality to decline. What would motivate the
other authors to allow such a decline? In this paper, we work with the basic
theoretical assumption that all authors are rational agents that seek to
maximize the quality of their work. This is why the unreasonableness of
allowing negative values of $\ln a$ in the fitting becomes even greater in the
``hard sciences'' in which the consensus on the methods and results (for
instance, theorems and proofs in computer science and mathematics; quantitative
measurements and models in the natural sciences) that constitute a paper is
clear. 

Nevertheless (anticipating our choice for measuring $q$ in the next
section), we note that while there is general support for the notion that the
``quality'' of a paper---when measured as the number of citations that it
accrues---benefits from the
work of additional authors \citep{figg06,bornmann07b,lokker08}, Waltman and van
Eck \citep{waltman15} find a very slight detrimental
effect on the citation counts of papers with three, four or five authors with
respect to papers authored by two authors (they are still cited substantially
more than papers by a single author). For six and more authors, an unequivocal
benefit is seen. Their analysis is based on an average of
field-normalized citation scores across all the disciplines in the Web of
Science database and seems to indicate, at first glance, that contrary to our
assumption additional authors may have a detrimental effect on the quality of a
joint paper.

While the results of \cite{waltman15} merit more careful scrutiny and an
analysis broken down by scientific fields, one possible reason for this
apparent average decline in quality with additional authors could be that larger
collaborations tend to split work over several different papers, a strategy
with a known benefit \citep{bornmann07b}, to a greater extent than the author
pair. In this case, the total citation count of that group of coauthors should
be the sum over their joint papers. We shall correct for this eventuality in
our analysis (\textit{vide infra}) by multiplying the author abilities by the
number of coauthored papers. However, if the motive were simply to minimize the
residuals in the fitting, a more malleable model with more fitting parameters
would be appropriate. Using such a strategy, the residuals can be made to
disappear completely but at the same time, the validity of the extracted
parameters is decreased. Nevertheless, at the express insistence of one of the
Reviewers, the analogous results of those given in the next section will are
provided in the Appendix.

\section{Illustrative real-world example}
For purposes of illustration, we take the variable $q_\alpha$ to correspond
to the number of citations of paper $\alpha$. We will then rank authors, not by
$a_i$ directly however, because that will give undue weight to the average
performance of an author, but rather by $n_i a_i$, where $n_i$ is the number
of papers to which author $i$ has contributed in the statistical sample. Like
this, we hope to cover both the ``breadth'' and ``depth'' of an author's
output. As the starting point for the iterative solution of
Eq.~(\ref{eq:quality}), we take the fractional number of citations per paper
for each author $i$. All numerical calculations were performed using the
\textsc{GNU Octave} \citep{octave} software, version 3.8.1.

\begin{figure}
\centering
\includegraphics*{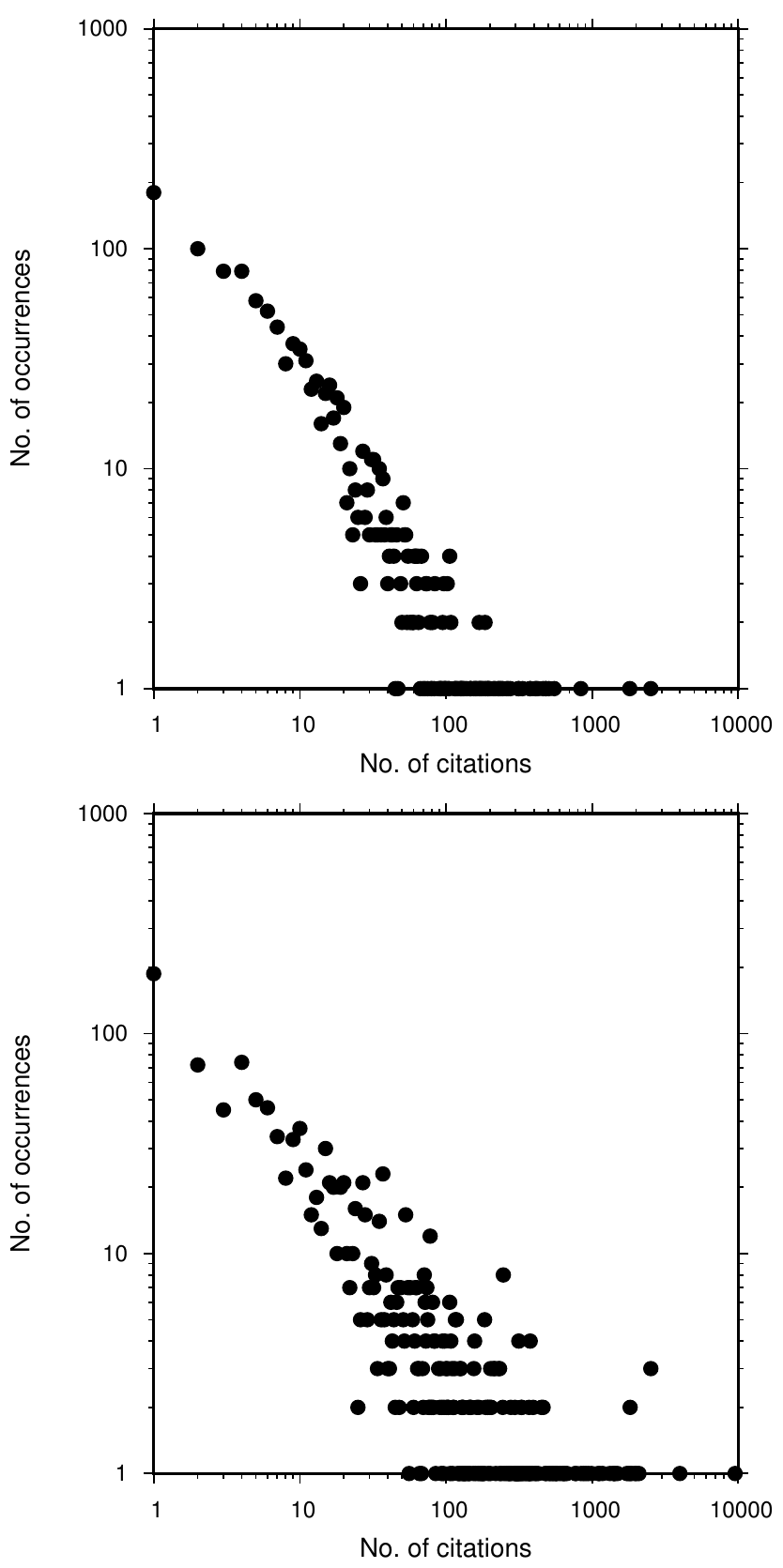}
\caption{(Top panel) Frequency distribution of paper citation counts in the
dataset. (Bottom panel) Frequency distribution
of author citation counts in the dataset.}
\label{fig:qdist}
\end{figure}

The statistical basis for this non-exhaustive study was obtained from the
CiteSeerX online database\footnote{\url{http://citeseerx.ist.psu.edu}, accessed
February, 2015.} by compiling the cited papers\footnote{We limit our study to
\emph{cited} papers, not out of theoretical necessity, but out of practical
convenience.} of renowned computer scientists Thomas H. Cormen\footnote{Search
query: \texttt{author:"thomas+h+cormen"}} and Charles E.
Leiserson\footnote{Search query: \texttt{author:"charles+e+leiserson"}} and
their immediate coauthors.\footnote{Search queries generated automatically by a
script on the same model as used for Cormen and Leiserson.} This search yielded
data for 1228 publications by a total of 1416 authors, after some manual
pruning for author name variations where ambiguity was not an issue
(\textit{e.~g.} ``James'' or ``Jim'') and also for some transcription errors in
the database (\textit{e.~g.} part of the title of the paper or author
information [affiliation, \textit{etc.}] contaminating an author name).
However, of these authors, 856 only appear on one paper each in the dataset and
were excluded from the regression analysis. This increases the robustness of
the results, as any statistical method is only reliable if there are repeated
occurrences in the dataset. No correction for ``inseparable coauthors''
(authors who invariably publish together) was made in the analysis, as such
groups are indistinguishable from a single author in output and citation data
and so cannot be mathematically disentangled. The frequency distributions for
the number of times a document or an author is cited are given in
Figure~\ref{fig:qdist} and are seen to exhibit the heavy tail typical of
citation distributions \citep{egghe98}. The statistical basis should be
sufficient for our purposes.

\begin{figure}
\centering
\includegraphics*{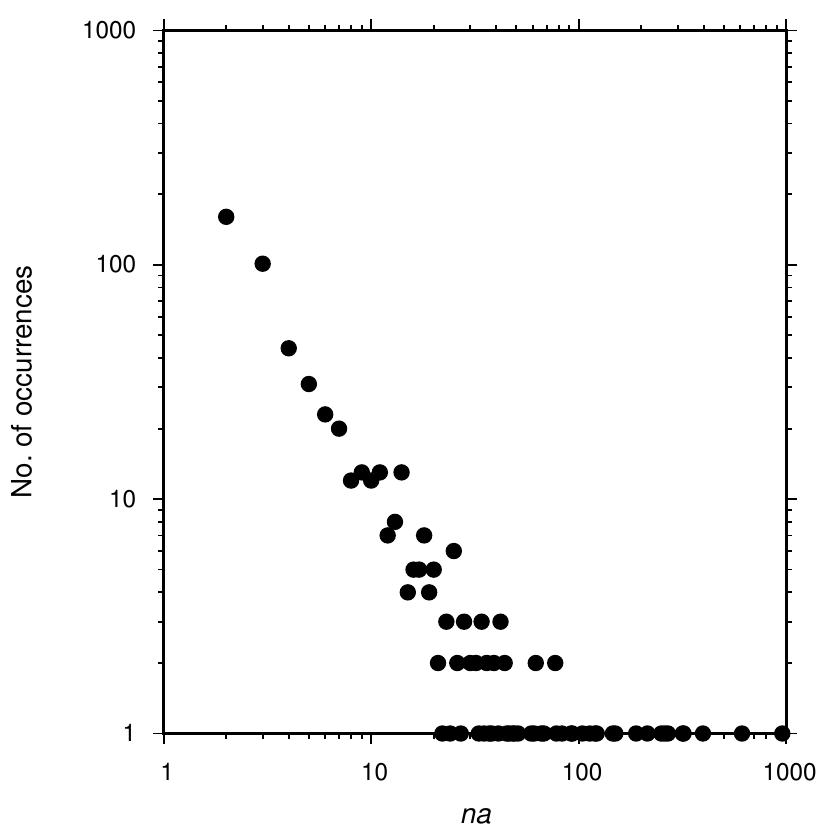}
\caption{The distribution of $n a$ (rounded to integer values) obtained from
the regression analysis.}
\label{fig:adist}
\end{figure}

A least-squares regression analysis was performed on the data to yield
a set of unique author abilities $\{a_i\}$. The values for $n_i a_i$ range
from $2$ to almost $1000$; the distribution is visualized in
Figure~\ref{fig:adist}. Evidently, the shape of the distribution of the $n
a$-values is reminiscent of those of the paper and author citations: most
authors are of ``ordinary'' ability and not easily distinguishable. The author
with the highest $n a$-value (and, incidentally, also the highest $a$-value)
in the dataset turns out to be renowned cryptologist Ronald L. Rivest (known
for the RSA cryptosystem). He is, however, not the most productive author in
the dataset, having fewer papers than David Kotz; he does, on the other hand,
have more citations than Kotz and so would rank higher also in most classical
rankings. The top-ten ranked authors are given in Table~\ref{tab:results} with
some bibliometric data from the dataset. The $n a$-ranking of the top ten
follows that of the total number of citations closely, but with some notable
exceptions: Sivan Toledo, David M. Nicol, Michael A. Bender and Robert D.
Blumofe all obtain a higher ranking under the $n a$-system than they would by
just counting total citations. Conversely, Satish Rao, Benny Chor and C. Greg
Plaxton obtain lower rankings under the $na$-system than they would by total
citations.

\begin{table}
\caption{Number of publications ($n$), $n a$-value, total and fractional number
of citations (with or without authors included that only appear once) as well
as the $h$-index ($h$) for the ten top-ranked authors in the dataset according
to $n a$-value. The value of $n a$, as well as that of the fractional citation
count, is rounded to the nearest integer. The Pearson correlation coefficient
between $na$ and the total citation count in this table is $r = 0.95$; between
$na$ and the fractional citation count in this table, it is $r = 0.94$ if
authors who only appear once are excluded and $r = 0.97$ if they are included.}
\label{tab:results}
\begin{center}
\begin{tabular}{l r r r r r r}
Author 				& $n$ & $n a$	& Citations	& Frac. cit.$^a$ & Frac.
cit.$^b$ & $h$ \\ \hline
Ronald L. Rivest	& 102 & 957   	& 9524 & 6531 &	3766 & 31		\\
David Kotz 			& 145 & 613   	& 3987 & 1900 &	1769 & 32	\\
Guy E. Blelloch 	& 71  & 398 	& 2006 & 997 & 929 & 23		 	\\
Robert D. Blumofe   & 13  & 321  	& 1780 & 963 & 600 & 11 \\
Michael A. Bender 	& 59  & 317    	& 1409 & 583 & 496 & 19 \\
David M. Nicol		& 68  & 270    	& 856  &	384	 & 319 & 17	\\
Satish Rao     		& 51  & 260    	& 1964 &	834	 & 664 & 22	\\
Sivan Toledo		& 60  & 251    	&  994 & 638 & 557 & 17	\\
Benny Chor 			& 41  & 215    	& 1824 & 793 & 590 & 18	\\
C. Greg Plaxton		& 46  & 189    	& 1857 & 771 & 595 & 17	\\	\hline
\multicolumn{6}{l}{$^a$ Authors who only appear once not counted.} \\
\multicolumn{6}{l}{$^b$ Authors who only appear once counted.}
\end{tabular}
\end{center}
\end{table}

The correlation between the integer citation count and the $n
a$-values apparent from Table~\ref{tab:results} is slightly stronger when the
fractional citation count \emph{including all authors} is substituted for the
integer one. This is actually a surprising result since the $na$ values are
calculated from a sample from which authors who only appear once have been
removed. The strong correlations are, nevertheless, somewhat attenuated when
the whole data sample is considered instead of only the most outstanding
authors: the Pearson correlation coefficient between $na$ and the total
citation counts for the whole dataset is $r = 0.89$; and between $na$ and the
fractional citation counts, it is either $r = 0.89$ (excluding authors who only
appear once) or $r = 0.92$ (including authors who only appear once). However,
perhaps more interesting for the purposes of author ranking is the rank
correlation. The Spearman rank correlation between the fractional citation
count and the $na$ values is $\rho = 0.79$ (when rounded to two decimal places,
the result is the same whether or not authors who only appear once in the
dataset are excluded or not from the denominator), which is slightly stronger
than the corresponding rank correlation of $\rho = 0.70$ with the total
citation count. 

\section{Concluding discussion}
While the $na$-ranks agree rather well with traditional measures of high-level
scientific productivity, contrary to the traditional approach which is purely
\textit{ad hoc}, the proposed model of this paper is based on the assumption
that the underlying scientific productivity is governed by a factor that can be
estimated from regression analysis. Arguably, the age-old adage: ``practice
makes perfect'' is likely to hold true to some extent also when performing
scientific research and writing scientific papers, but in the interest of
keeping the unknown parameters to a minimum, we have not considered this effect
in our model. Nevertheless, the results support the view that fractional
citation counting is a fair way to distribute credit, at least within the
computer science field. In line with this finding, it is important to stress
that the strong rank correlation between citations (fractional or otherwise)
notwithstanding, the idea in this paper is not to introduce a more
``expensive'' method to calculate the citation ranks. It is only the
differences with respect to the traditional ranking that are interesting,
because they show precisely the extent to which there is a need to step away
from the simplified author ranking for purposes of promotion and funding.  

It is interesting to compare the proposed method with that of \cite{tol11}, seeing as it is the one with which it shares
the most of the undergirding philosophy. Contrary to \cite{tol11},
there is no need to assume any form for the citation distribution. Since 
\cite{tol11}, implicitly at least, assumes an unvarying distribution for
each author,\footnote{The distributions that \cite{tol11} considers
change through the iterations used to solve the model, but the converged
result is a function, like the $a$-value, only of the bibliographic record and
does not change for one and the same author from one paper to the next.}
his model is also based on the concept of an unchanging, inherent ``author
ability'' that is used to produce cited papers. The proposed method is hence
seen to be more general in its assumptions. For instance, the ``ability'' to
publish pages of scientific output could just as well be the underlying
variable that we wish to extract statistically; \textit{i.~e.}, the
bibliometric indicator could be the number of pages per paper instead of
citations. The idea is that one first identifies a measure of quality ($q$) for
the individual paper, and then proceeds to analyze the underlying distribution
of the authors' abilities ($a$).

Note that one of the basic ideas in the Shen-Barab\'asi \citep{shen14}
approach---to distinguish coauthor disciplines through their
degree of ``cocitedness'' with other papers (essentially distinguishing
scientific disciplines by the sets of papers that cite a particular paper)---is
easily adapted to the current algorithm. One needs simply to redefine the
quantity $q$ accordingly by, for instance, defining $q_\alpha$ to be a weighted
sum of citations, in which the weight of a citation to paper $\alpha$ from
paper $\beta$ is determined by the ``cocitation strength'' \citep{shen14}
between papers $\alpha$ and $\beta$: \textit{i.~e.}, the number of papers
citing both $\alpha$ and $\beta$. This is an interesting avenue for further
development.

Finally, I stress once more that in some extreme cases, individual author
abilities cannot be distinguished even in principle. This occurs, for instance,
when two authors are ``inseparable coauthors'', and the one never publishes a
paper without the other. This problem is, however, endemic to the whole domain
of citation analysis and becomes less of an issue in practice as the seniority
of an author increases.

\section*{Acknowledgment}
I thank the anonymous referees for helpful suggestions.

\appendix
\section{Regression with ``destructive authors'' in the dataset}
If we relax the requirement that $\ln a_i \geq 0$ for any author $i$, we assume
that said author $i$ is a ``destructive force'' which unbeknownst to his
coauthors and himself sabotages the paper they produce. For completeness, we
provide the resulting ``top ten'' authors using this assumption in
Table~\ref{tab:destructive}. This provides an indirect measure of the
robustness of the method.

\begin{table}
\caption{Number of publications ($n$), $n a$-value and total number
of citations for the ten top-ranked authors in the dataset according
to $n a$-value \emph{when authors are allowed to have $a$-values less than
unity in the statistical fitting}. The value of $n a$ is rounded to the nearest
integer.}
\label{tab:destructive}
\begin{center}
\begin{tabular}{l r r r}
Author 				& $n$ & $n a$	& Citations	\\ \hline
Ronald L. Rivest	& 102 & 1272  	& 9524 \\
David Kotz 			& 145 & 957   	& 3987 	\\
Guy E. Blelloch 	& 71  & 776 	& 2006 	\\
James Demmel        & 7   & 725  	& 631  \\
Marc Moreno Maza	& 45  & 618    	& 514 	\\	
Michael A. Bender	& 59  & 428		& 1409 \\
Sivan Toledo		& 60  & 374		& 994 \\
David M. Nicol		& 68  & 339		& 856 \\
Anastassia Ailamaki & 4	  & 337		& 116 \\ 
Robert D. Blumofe	& 13  & 333		& 1780 \\ \hline
\end{tabular}
\end{center}
\end{table}

Like before the top two spots are still claimed by Rivest and Kotz (while now
their $n a$-values are higher for obvious reasons). With the exception of
Demmel, Maza and Ailamaki, all of the top ten names appear also in
Table~\ref{tab:results}, indicating only a slight reordering. The rank
correlations between the $n a$-values and the number of citations are $\rho =
0.67$ (total), $\rho = 0.73$ (fractional with all authors) and $\rho = 0.74$
(fractional excluding one-time authors) in the whole dataset. The corresponing
Pearson correlation coefficients are $r = 0.78$, $r = 0.81$ and $r = 0.78$,
respectively. Thus, even with this ``unphysical'' assumption, we see a stronger
correlation with fractional citation counts than with the integer one.  


\end{document}